\documentclass[twocolumn,showpacs,preprintnumbers,superscriptaddress,prb,floatfix,aps,10pt]{revtex4-1}

\usepackage{hyperref}
\hypersetup{
    bookmarks=true,         % show bookmarks bar?
    unicode=true,           % non-Latin characters in Acrobat’s bookmarks
    pdftoolbar=false,       % show Acrobat’s toolbar?
    pdfmenubar=true,        % show Acrobat’s menu?
    pdffitwindow=true,      % page fit to window when opened
    pdftitle={My title},    % title
    pdfauthor={Author},     % author
    pdfsubject={Subject},   % subject of the document
    pdfnewwindow=true,      % links in new window
    pdfkeywords={keywords}, % list of keywords
    colorlinks=true,        % false: boxed links; true: colored links
    linkcolor=blue,         % color of internal links
    citecolor=blue,         % color of links to bibliography
    filecolor=black,        % color of file links
    urlcolor=blue           % color of external links
  }

\usepackage{amsmath,amssymb}
\usepackage{graphicx,textcomp}
\usepackage{subfigure}
\usepackage{color}
\usepackage{todonotes}
\usepackage[titletoc,toc,title]{appendix}
\usepackage{xcolor}
\usepackage{todonotes}
\usepackage[utf8]{inputenc}
\definecolor{bjorn}{RGB}{255,0,0}
\definecolor{davide}{RGB}{0,255,0}

\newcommand{\angstrom}{\text{\normalfont\AA}}

\begin{document}

\title{Lattice relaxations in disordered Fe-based materials in the paramagnetic state from first principles}
\author{Davide Gambino}
\email{davide.gambino@liu.se}
\affiliation{Department of Physics, Chemistry, and Biology (IFM), Link\"{o}ping University, SE-581 83, Link\"oping, Sweden.}

\author{Bj\"{o}rn Alling}
\affiliation{Department of Physics, Chemistry, and Biology (IFM), Link\"{o}ping University, SE-581 83, Link\"oping, Sweden.}
\affiliation{Max-Planck-Institut f\"ur Eisenforschung GmbH, D-402 37 D\"usseldorf, Germany.}

\date{\today}

\begin{abstract}

The  first-principles calculation of many material properties, in particular related to defects and disorder, starts with the relaxation of the atomic positions of the system under investigation. 
This procedure is routine for nonmagnetic and magnetically ordered materials. 
However, when it comes to magnetically disordered systems, in particular the paramagnetic phase of magnetic materials, it is not clear how the relaxation procedure should be performed or which geometry should be used. 
Here we propose a method for the structural relaxation of magnetic materials in the paramagnetic regime, in an adiabatic fast-magnetism approximation within the disordered local moment (DLM) picture in the framework of density functional theory (DFT). 
The method is straight forward to implement using any $ab$ $initio$ code that allows for structural relaxations.
We illustrate the importance of considering the disordered magnetic state during lattice relaxations by calculating formation energies and geometries for an Fe vacancy and C insterstitial atom in  bcc Fe as well as bcc Fe$_{1-x}$Cr$_x$ random alloys in the paramagnetic state. 
In the vacancy case, the nearest neighbors to the vacancy relax towards the vacancy of 0.16 \angstrom (-5\% of the ideal bcc nearest neighbor distance), 
which is twice as large as the relaxation in the ferromagnetic case. The vacancy formation energy calculated in the DLM state on these positions is 1.60 eV, which corresponds to a reduction of about 0.1 eV compared to 
the formation energy calculated using DLM but on ferromagnetic-relaxed positions. The carbon interstitial formation energy is found to be 0.41 eV when the DLM relaxed positions are used, as compared to 0.59 eV when the FM-relaxed positions are employed. 
For bcc Fe$_{0.5}$Cr$_{0.5}$ alloys, the mixing enthalpy is reduced by 5 meV/atom, or about 10\%, when the DLM state relaxation is considered, as compared to positions relaxed in the ferromagnetic state. 

\end{abstract}
\maketitle

\section{Introduction\label{Introduction}}

%Relaxation in NM and FM materials

The study of defects in crystals from first principles has for a long time assisted the community with qualitative understanding and explanations of findings related to phase stability \cite{Igor_phasestability_defects}, diffusion \cite{diffusion_properties}, electronic \cite{Neugebauer_el_relax_defects} and optical \cite{optical_properties} properties. Nowadays, advanced calculations are approaching quantitative predictive accuracy \cite{PRX_Glensk,RevModPhys_Freysoldt}. 
Many investigations of structural disorder in the form of vacancies, interstitial atoms and substitutional alloys, are present in the literature: consistently, local lattice relaxations around the defects or throughout the crystal for alloys, are known to affect the energetics of the system, and sometimes explicitly its properties. 
Algorithms to perform lattice relaxations are commonly implemented in many softwares for first principles calculations,
and the main working principle consists in moving the atoms according to the forces to which they are subject, in order to obtain the equilibrium structure of the system.
These techniques work very well for nonmagnetic (NM), ferromagnetic (FM), or antiferromagnetic (AFM) materials; however, when it comes to magnetic materials in the high temperature paramagnetic phase (PM), there is a methodological gap.

%PM state: DLM

This gap is due to the difficulties in the treatment of the PM phase from first-principles. 
In the past, the PM phase has sometimes been modelled as a NM phase, but this approach leads to wrong results if applied straightforwardly\cite{rivadulla,CrNright_Alling} in the framework of density functional theory (DFT). 
The reason is that,  above the critical temperature, while the system does not show any macroscopic magnetic field nor long range magnetic order, the local spin polarization of the electron density around the atoms is typically retained and do influence the behavior of the material.
In order to model appropriately this state of matter,  advanced methods need to be applied, and one of the possibilities is the disordered local moment (DLM) approach \cite{Hubbard_I,*Hubbard_II,*Hubbard_III,Hasegawa_I,*Hasegawa_II,DLM_Gyorffy}.

The applicability of the DLM model is motivated by the fact that the real magnetization density in many magnetic materials can be described in terms of quite robust magnetic moments localized close to the atoms. In this case, its PM phase can be viewed as a disordered distribution
of such local moments without long-range order. In DLM simulations, the direction of the moments is assigned randomly, i.e. the correlation function between the moments is $\approx$ 0. 
Since the model neglects possible short-range-order, the PM state described corresponds in principle to the high-temperature limit $\frac{J_{ij}}{k_B T}\to 0$, where $J_{ij}$ is the strongest magnetic interaction in the system. 
The DLM model was originally implemented in a coherent potential approximation (CPA) framework \cite{DLM_Gyorffy}, but has since been used also in supercell approaches \cite{CrNright_Alling,DLM_MSM_Alling}. 
The DLM picture is easily implemented in DFT calculations; it can be combined with techniques that include strong electron correlation effects (such as the LDA+U scheme)\cite{DLM_MSM_Alling}, and can be employed both with collinear and noncollinear magnetic moments.

A further complication is that the PM state is also disordered in time, meaning that the direction of the moments change quickly compared to the jump rates of atoms and defects, and often even compared to the vibrational frequencies of the atoms. 
Thus, one must be careful when employing a static DLM model for the PM state, in particular when the atoms are allowed to move away from high-symmetry points like in the case of lattice relaxations or during molecular dynamics (MD). 

For the latter case, a particular method, DLM-MD, has been developed \cite{DLMMD_Steneteg,DLM_CrAlN_Alling,DLMMD_Shulumba,DLMMD_Mozafari} were the magnetic state is rapidly changed between different disordered configurations during the MD. 
Recently, an even more accurate model combining atomistic spin dynamics with ab initio molecular dynamics (ASD-AIMD)\cite{ASD-AIMD_Irina} has been suggested in order to include lattice vibrations into the description of the PM state. 
However, very little attention has been given to static lattice relaxations around defects or in alloys \cite{DLM_CrAlN_Alling}. 

%Common practice and DMFT relaxation

Because of the lack of a method that performs lattice relaxations for the paramagnetic state, the atomic positions obtained from a relaxation performed with FM moments are commonly employed \cite{SW_Ruban,SW_1vFe_Asta,DLM_1vFe_Korzhavyi,DMFT_1vFe}.
If we neglect spin-orbit coupling, the FM state do have the same lattice symmetry as the PM state, but it is known that the interatomic bond strengths can differ substantially \cite{SSA_bccFe} putting doubt on the reliability of this approach.
Nevertheless, formation energies of vacancy \cite{SW_Ruban,SW_1vFe_Asta,DLM_1vFe_Korzhavyi} and many substitutional defects \cite{substitutionals_Fe_KR1,substitutionals_Fe_KR2} have been calculated for Fe in the paramagnetic phase, but relaxed in the FM state.
%In the DLM approach, formation energies in the PM state are calculated as an average over several disordered magnetic configurations, or in some cases using a special quasirandom structure, using the FM-relaxed positions.
It has also been suggested, that a complete relaxation in each frozen DLM magnetic configuration can be performed and that the artificial extra relaxation can cancel out, at least when studying mixing enthalpies or defect formation energetics \cite{DLM_CrAlN_Alling,C_int_fccFe_Igor}.
Recently, the vacancy formation energy in PM bcc Fe has been calculated also by means of DFT plus dynamical mean field theory (LDA+DMFT) relaxing the first two shells of atoms neighboring to the vacancy by a manual energy-minimization procedure \cite{DMFT_1vFe}. 
The results do show a non-negligible relaxation energy compared to the FM relaxed geometry.
Nonetheless, a robust demonstration of force calculations in DMFT would be needed before more complex relaxation problems could be adressed with this computationally demanding approach.

For this reason, in the present article we propose a method, based on the DLM approach within DFT, which allows to obtain atomic structures relaxed in the PM phase. The method can be performed with any first principles software which allows for the reliable calculation of interatomic forces.
The key idea of the method is that the atoms are partially allowed to relax according to different DLM states in sequence: as a result, fluctuations in forces originating in the particular magnetic states are averaged on-the-fly during the relaxation.
The symmetry of the underlying lattice is imposed when applicable (e.g., not in the case of a substitutional alloy), disregarding the disorder of the magnetic state.
The procedure is iterated until a steady displacement of the atoms from the initial positions is achieved. 
The equilibrium positions are finally obtained averaging over several atomic configurations in the steady displacement regime. 

We initially test the method on PM bcc Fe with a single Fe vacancy, and we show the importance of lattice relaxations according to the relevant magnetic state, both with and without imposition of symmetry; 
then, we we perform the lattice relaxation of defect free PM bcc Fe without imposition of symmetry, in order to show the level of accuracy achievable. 
We finally apply the method to the case of C interstitial in octahedral position in PM bcc Fe, and PM bcc Fe$_{1-x}$Cr$_{x}$  alloys ($x$ = 0.25, 0.5, 0.75). 
All of these cases are of high relevance for the development of steels and motivate an attempt to increase the quantitative accuracy of simulating their energetics.
%Following, we apply the relaxation method to the study of the Cr vacancy in PM B1 CrN, a system with strong electronic correlation effects and an antiferromagnetic (AFM) ground state, which makes it more difficult to relax consistently with standard techniques.

The paper is organized as follows. In Sec. \ref{CompDet}, the computational details are described. 
Sec. \ref{accuracyDLMcalculations} deals with the description of the PM state in terms of the DLM model, together with a comparison between results from collinear and noncollinear DLM calculations. 
In Sec. \ref{Method}, the present method is outlined following the illustrative case of the relaxation of bcc Fe with a vacancy, and the uncertainty on the relaxed positions for symmetryless relaxation is discussed with the case of defect free bcc Fe.
The results on the vacancy and the C interstitial in octahedral positions in PM bcc Fe are presented in Sec. \ref{1vbccFe} and \ref{octCbccFe} , respectively. The intermetallic alloy bcc Fe$_{1-x}$Cr$_{x}$ is discussed in Sec. \ref{bccFeCr}. 
In Sec. \ref{Conclusions}, we draw the conclusions of the present work.

\section{Theoretical Methods \label{TheoMethod}}

\subsection{Computational details \label{CompDet}}

The first principles calculations are carried out in the framework of DFT using projector augmented wave (PAW) potentials \cite{PAW_Blochl,PAW_vasp} as implemented in the Vienna $ab$ $initio$ simulation package (VASP) \cite{vasp_I,*vasp_II} and a plane wave energy cutoff of 400 eV. 
The accuracy of the self-consistent calculations is set to  $10^{-4}$ eV/supercell.
The PBE exchange and correlation functional \cite{PBE} is employed for all calculations.
We have chosen to perform all the calculations with the theoretical 0 K lattice parameter since we want to prove the importance of lattice relaxations on the energetics of the system, rather than aiming at accurate results mimicking a particular temperature.

Calculations of bcc Fe with defects are performed using supercells composed of 3x3x3 bcc conventional cells (54 Fe lattice sites) with lattice parameter of 2.84 \angstrom, while sampling the first Brillouin zone with a 3x3x3 k-points mesh according to the Monkhorst-Pack scheme \cite{MPscheme}.
%Regarding CrN, which has a B1 rock-salt structure, we employ a 2x2x2 supercell (theoretical lattice parameter = 4.13 \angstrom) and a 3x3x3 Monkhorst-Pack k-points mesh.
For bcc Fe$_{1-x}$Cr$_{x}$ alloys, the considered supercell consists of 4x4x4 bcc primitive cells, for a total amount of 64 atomic sites per supercell, with atoms distributed in a special quasirandom structure (SQS)\cite{SQS}. 
In this case, the relaxation of the structures in the DLM state was performed sampling the Brillouin zone with a Gamma-centered 2x2x2 k-points mesh, whereas the final DLM calculations on the obtained geometry were performed on a 3x3x3 Monkhorst-Pack k-points mesh, as in the other cases.
The theoretical lattice parameter for each composition of the alloy is derived from 0 K FM calculations.

In order to simulate the PM state, we employ the DLM method using a supercell approach. The DLM calculations are carried out using noncollinear magnetic moments, constrained in the selected direction using the method developed by Ma and Dudarev \cite{ConstrMag_Dudarev}. 
For bcc Fe, the constraining parameter $\lambda$ was set to 10, whereas in the case of bcc Fe$_{1-x}$Cr$_{x}$ alloys we started with $\lambda=1$ and increased the parameter to $\lambda=10$ to facilitate convergence of the electronic calculations. 

We show the applicability of the present method also with collinear calculations in the case of defect free bcc Fe.
The symmetry analysis is carried out with the Phonopy \cite{phonopy} package.

The vacancy formation energy $E_{\text{1v}}^{f}$ is calculated according to:
\begin{equation}
E_{\text{1v}}^{f}=E_{\text{1v}}-(N-1) E_0(\text{Fe}),
\end{equation}
where $E_{\text{1v}}$ is the energy of a supercell with $N$ lattice sites, of which $N-1$ are occupied by Fe atoms and one is vacant, and $E_0(\text{Fe})$ is the energy of one Fe atom in defect free bcc Fe in the relevant magnetic state (FM or DLM).
The C interstitial formation energy $E_{\text{octC}}^f$ is calculated with equation:
\begin{equation}
E_{\text{octC}}^{f}=E_{\text{octC}}-N E_0(\text{Fe})-E_0(\text{C}),
\end{equation}
where $E_{\text{octC}}$ is the energy of a supercell with $N$ Fe atoms and one C atom in octahedral position, $E_0(\text{Fe})$  is the energy of one Fe atom as described above, and $E_0(\text{C})$ is the energy of one C atom in diamond.
We chose to use diamond as reference state for C in order to avoid the technical complexities in calculations of graphite (which is the ground state of C) due to van der Waals interactions \cite{Marzari_graphite,graphite_functionals}.
The mixing enthalpy $\Delta E_{mix}$ of the random alloy at concentration $x$ is calculated as:
\begin{equation}
\Delta E_{mix}(\text{Fe}_{1-x}\text{Cr}_x)=E(\text{Fe}_{1-x}\text{Cr}_x)-(1-x)E(\text{Fe}) - xE(\text{Cr})
\end{equation}
where $E(\text{Fe}_{1-x}\text{Cr}_x)$ is the energy of the alloy at concentration $x$, and $E(\text{Fe})$ and $E(\text{Cr})$ are the energies of the reference states, which are bcc Fe in the relevant magnetic state, and nonmagnetic Cr. 

In this work we do not calculate free energies. 
Since the DLM state is the state with maximally disordered magnetic moments, its magnetic entropy will also be maximum. 
However, when comparing calculations on different atomic positions or calculations with and without defects with the same number of atoms in the same magnetic state, this magnetic entropy, within this DLM approximation, will not contribute to the free energy difference because it cancels out.

\subsection{Description of the paramagnetic state with collinear and noncollinear DLM models\label{accuracyDLMcalculations}}

In order to correctly model the PM phase of the system under investigation, we employ a magnetic sampling method (MSM) \cite{DLM_MSM_Alling}, 
which consists in  performing static calculations on fixed geometries using a large set of DLM configurations of the magnetic moments obtained with random number generation; 
the property of interest is then obtained as an arithmetic average over all the employed configurations. 
The error on the mean value of the property $P$ ($\sigma_{\bar{P}}$)  is calculated as the standard error, i.e. $\pm 2\sigma_{\bar{P}}=2\sigma_{P}/\surd N$, with confidence interval of 95\%, where $\sigma_{P}$ is the standard deviation of the results associated to the $N$ DLM configurations. 

\begin{figure}
\begin{center}
\includegraphics[width=\columnwidth]{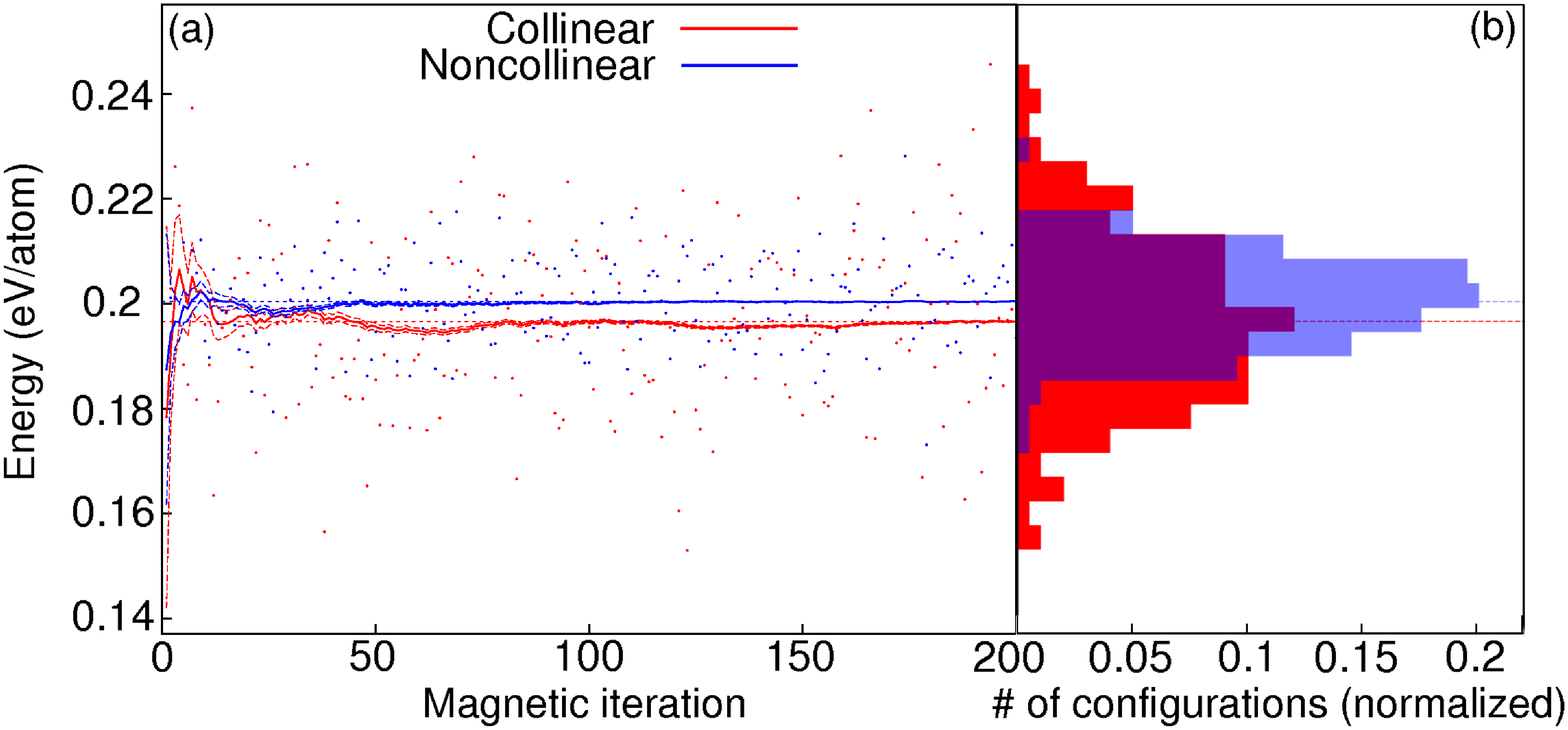}
\caption{(Color online) (a) Individual and accumulated average of energies calculated on the ideal lattice positions of defect free bcc Fe with collinear (red) and noncollinear (blue) magnetic moments. 200 random MSM configurations are employed here. 
The zero is taken as the calculated energy of FM bcc Fe. The dashed lines following the accumulated averages denote the standard error, whereas the dots are the calculated energies for a given magnetic iteration.
(b) Histogram of the energies, with areas of collinear and noncollinear samples normalized to one. Average energies are marked with dashed horizontal lines \label{histogram_E_dfbccFe}}
\end{center}
\end{figure}

\begin{figure}
\begin{center}
\includegraphics[width=\columnwidth]{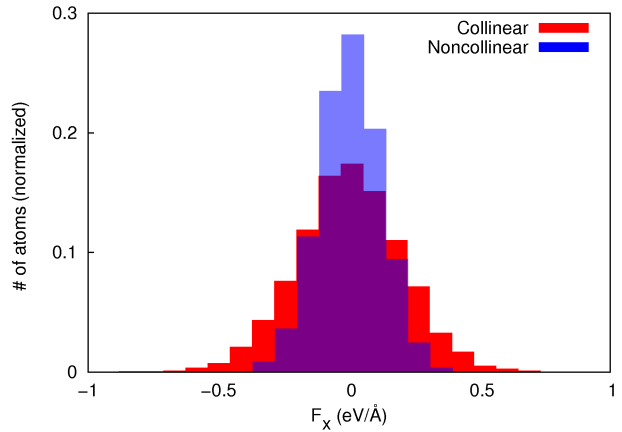}
\caption{Histogram of the x cartesian component of the forces calculated on the ideal positions of defect free bcc Fe with collinear (red) and noncollinear (blue) magnetic moments. Areas are normalized to one. \label{histogram_Fx_dfbccFe}}
\end{center}
\end{figure}

We carry out MSM calculations of defect free bcc Fe on the ideal lattice positions as a preliminary test in order to assess the statistical accuracy of the results. $N=200$ configurations are employed in this case, and both collinear and noncollinear magnetic moments are tested. 
The calculated energies and forces for this system are represented in Fig. \ref{histogram_E_dfbccFe} and \ref{histogram_Fx_dfbccFe}, respectively.
In the DLM framework, the energy $E_i$ and forces $\mathbf{F}_{i}$ of the system in a particular MSM configuration $i$  can be decomposed into two components, the first due to the average DLM phase, and the second due to the specific configuration employed, i.e.:
\begin{equation}
\begin{split}
E_{i}=E_{\text{DLM}}+\tilde{E}_{i},
\\
\mathbf{F_{i}}=\mathbf{F}_{\text{DLM}}+\mathbf{f}_{i},
\end{split}
\end{equation}
where, in the limit of infinite MSM configurations, $\langle\tilde{E}_{i}\rangle=0$ and $\langle \mathbf{f}_{i}\rangle= \bar{0}$, so that, by averaging the energy and forces of the system calculated with many different MSM states, the corresponding values of the DLM phase are retrieved as 
$\langle E_{i}\rangle \rightarrow E_{\text{DLM}}$ and $\langle \mathbf{F}_{i}\rangle \rightarrow \mathbf{F}_{\text{DLM}}$.
The same thing is true for pair-correlation functions, which explains why the MSM method yield the same results, within statistical error bars, as an SQS description of DLM \cite{DLM_MSM_Alling}.

The average energy of DLM defect free bcc Fe calculated with noncollinear magnetic moments, Fig. \ref{histogram_E_dfbccFe}, taking the energy of FM bcc Fe as reference, is $200 \pm 1$ meV/atom, whereas for collinear calculations the result is $197 \pm 2$ meV/atom, 
showing a small but distinguishable difference between the two different representations.
The energy converges to its average value within 1 meV in 40 configurations for the noncollinear case, as well as its confidence interval (dashed line around the cumulative average). 
For collinear magnetic moments, the energies of the individual DLM configurations are more spread and convergence is slower.
The forces on a specific atom in a particular configuration can be as strong as 0.5 eV/\angstrom, as shown in the histogram in Fig. \ref{histogram_Fx_dfbccFe} (here only one cartesian component of the force is considered); 
however, the distribution of the forces is centered at 0 as expected from conservation of center of mass (with standard error of 0.002 eV/\angstrom). It is clear from here that the distribution of the forces in the collinear case is larger than in the noncollinear one.
This means that in a relaxation based on collinear forces, fluctuations would be larger than in the noncollinear case and a larger degree of uncertainty would be introduced in the process. 

From these considerations, one can see that on average the forces on an atom are null when symmetry considerations can be applied; 
nevertheless, if one consider the atoms in the supercell as not equivalent, and consequently the averages are performed separately on each atom over the 200 different MSM configurations, the resulting mean forces can be as high as 0.02 eV/$\angstrom$  per component.
This is a statistical problem, so that in order to obtain 0 forces one needs either to use a very large amount of configurations, or to impose the symmetry of the underlying lattice to the system to recover the results shown in Fig. \ref{histogram_Fx_dfbccFe}.
Such symmetry imposition is easily done for a perfect lattice by averaging over all the atoms in the supercell, but requires a more careful analysis for a system with defects, and is not applicable to the case of a substitutional random alloy where every atom is formally unique. 

\subsection{Relaxation Method \label{Method}}

\begin{figure}
\centering
\subfigure{\label{trajectory_1vbccFe.a} \includegraphics[width=0.4\textwidth]{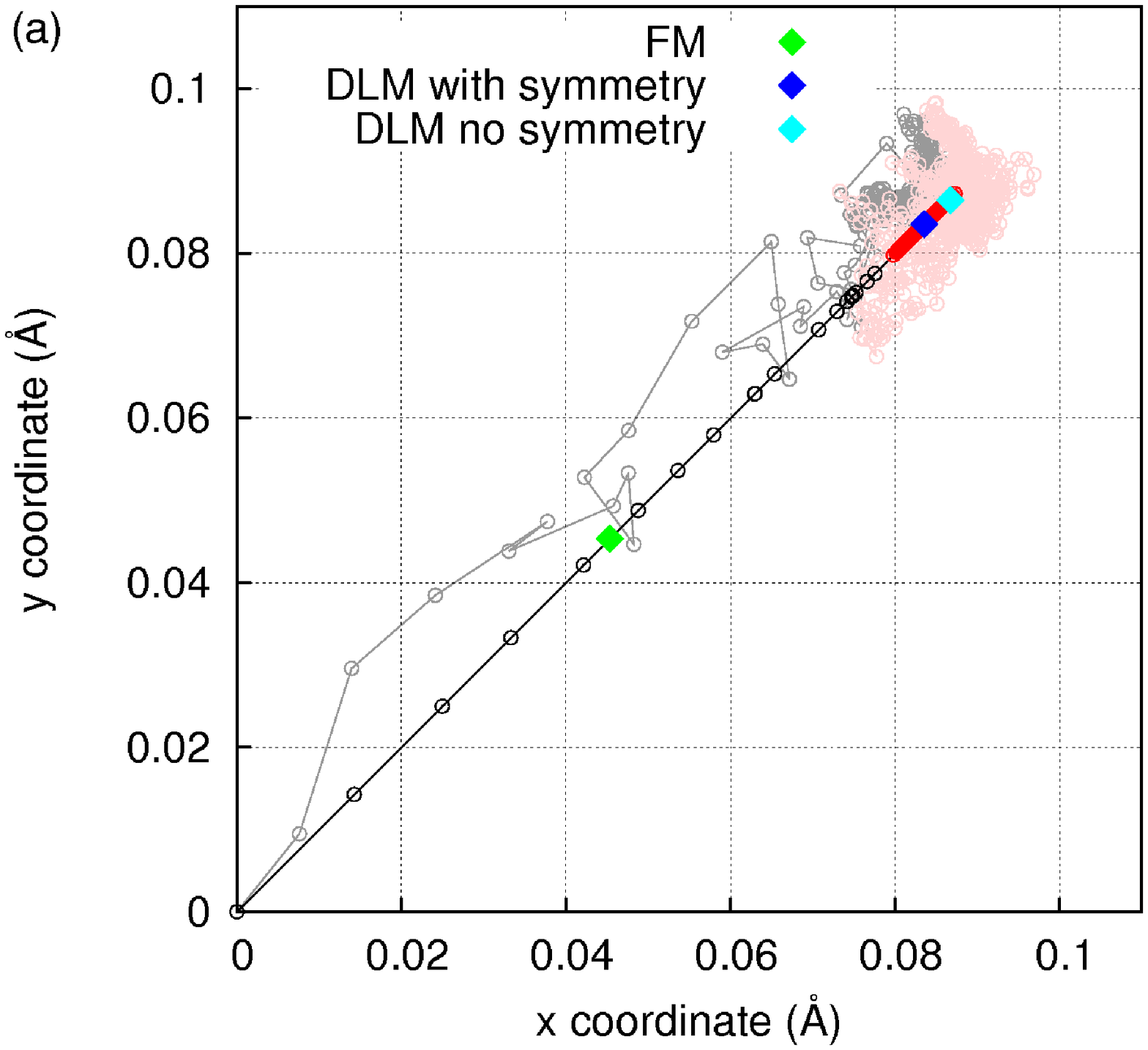}}
\subfigure{\label{trajectory_1vbccFe.b} \includegraphics[width=\columnwidth]{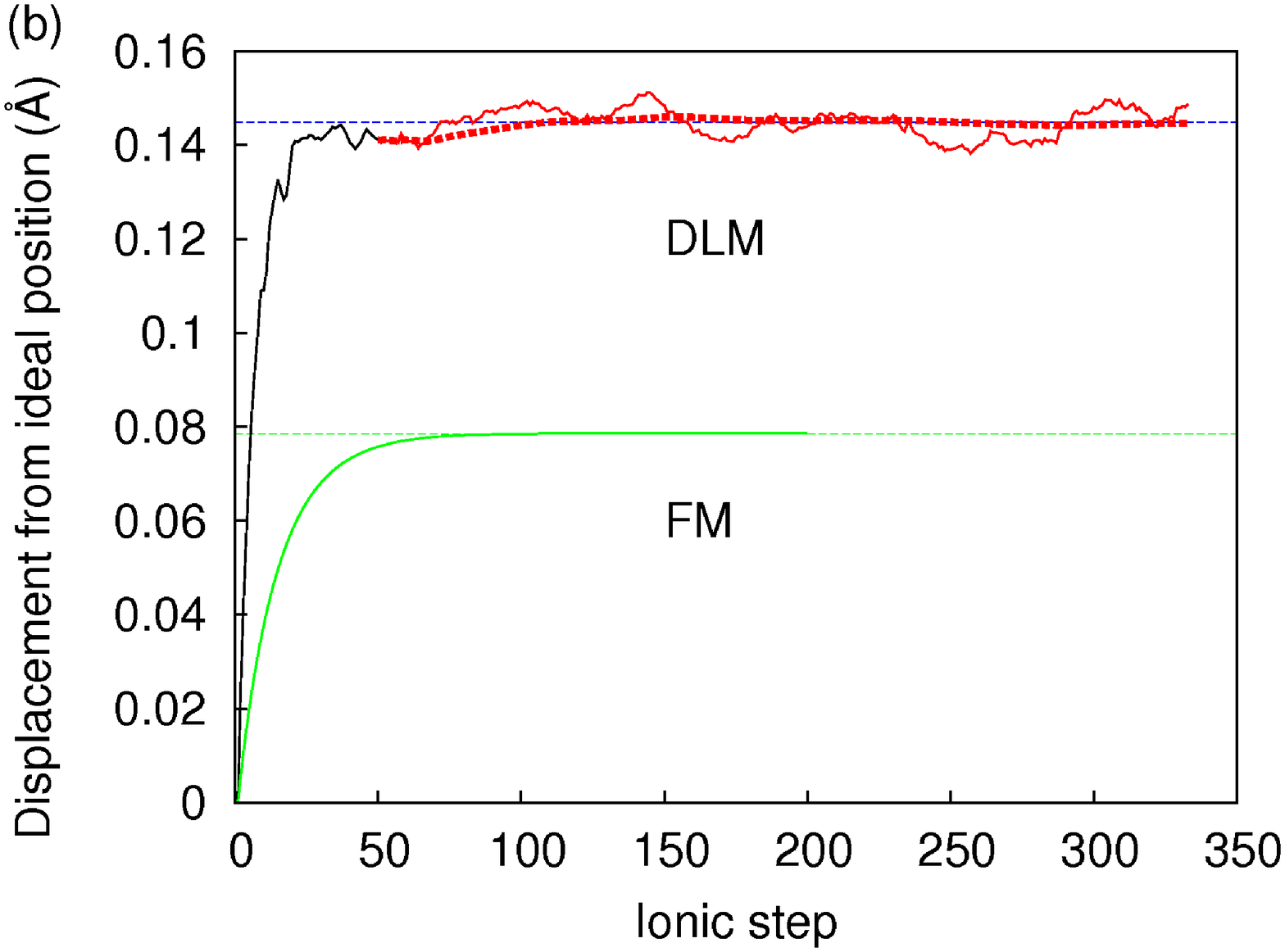}}
\caption{(a) x-y projection of the trajectory of one atom being nearest neighbor to the vacancy in the DLM-relaxation of bcc Fe with one vacancy, starting from its ideal lattice position in the origin. Solid and semi-transparent lines and points correspond to relaxation with and without imposition of symmetry, respectively. 
The red region consists of the positions averaged to get the equilibrium position. The final position using FM, DLM with imposition of symmetry, and DLM without imposition of symmetry are marked with green, blue, and light blue diamonds, respectively.
(b) Displacement from ideal lattice position as a function of the relaxation step. The black and red lines correspond to the relaxation with imposition of symmetry in part (a), and the blue dashed line indicates the displacement of the equilibrium DLM-relaxed position.
For comparison, the green lines illustrates the same relaxation procedure but for the FM magnetic state.
\label{trajectory_1vbccFe}}
\end{figure}

In Fig. \ref{trajectory_1vbccFe.a} an example of the relaxation procedure in the case of bcc Fe with one Fe vacancy is shown, both with and without imposition of symmetry (solid and semi-transparent lines and points, respectively);
 here, the xy-projection of the trajectory of an atom nearest neighbor to the vacancy (which is in a $\langle 111 \rangle$  direction in the bcc structure) is displayed for illustrative clarity. 

The relaxation procedure starts from an initial geometry of the system, being it the ideal crystal structure or some structure pre-relaxed by other means (e.g. FM-relaxed); in this example, we start from the ideal lattice positions (black empty circle at the origin, top figure).
A noncollinear MSM state is chosen randomly and a static calculation is performed on the initial geometry; from here, the forces are calculated and symmetrized, and the atoms are moved (first black empty circle on the right of the origin in Fig. \ref{trajectory_1vbccFe.a}), according to:
\begin{equation}
\mathbf{x}_{j+1}=\mathbf{x}_{j}+\alpha \mathbf{F}_{j},
\label{eq_xj+1}
\end{equation}
where $j$ indicates the current step of the relaxation, $\mathbf{x}$ and $\mathbf{F}$ are positions and forces, respectively, and $\alpha$ is a rescaling factor that will be further discussed later in this section.
The symmetrization of the forces is performed by analyzing the symmetry of the initial atomic geometry (in this example, the ideal bcc crystal structure with one vacancy), and storing the symmetry operations that connect equivalent atoms: 
in this way, the lattice sites in the supercell are mapped onto  a smaller set of sites that are symmetrically independent from each other.
When the static calculation is performed, the forces acting on each atom are projected onto the atom at the corresponding independent lattice site by application of the relevant symmetry operations,
 and averaged in order to obtain one force per site that respect the symmetry of the supercell. The symmetrized forces are then projected back to the equivalent atoms with the inverse symmetry operations.
After the update of the positions, another random MSM configuration of the magnetic moments is taken, and the same procedure is repeated: the forces are calculated and symmetrized, and the atoms are moved according to Eq. \ref{eq_xj+1}. 
Iterating this scheme, the trajectory indicated by the black line and circles in Fig. \ref{trajectory_1vbccFe.a} is obtained. In Fig. \ref{trajectory_1vbccFe.b}, the displacement of this atom from the ideal position during the relaxation is shown (all the components of the displacements are included here).
After a first transient in which the displacement increases almost monotonously (black solid line in Fig. \ref{trajectory_1vbccFe.b}), the relaxation comes to a stage where the displacement is fluctuating around some mean value (red solid line). 
This stage corresponds to the red area in Fig. \ref{trajectory_1vbccFe.a}, which consists of the positions that are averaged in order to obtain the PM equilibrium positions (blue solid diamond).
As a comparison, the FM-relaxed position is also shown in Fig. \ref{trajectory_1vbccFe.a} (green solid diamond), which is at a distance from the ideal structure position denoted by the green dashed line in Fig. \ref{trajectory_1vbccFe.b}.

The procedure is terminated when the accumulated average of positions (red dashed line in Fig. \ref{trajectory_1vbccFe.b}) in the roughly-constant displacement regime, where the atoms span stochastically the region around equilibrium thanks to the randomness of the magnetic moments, converges.
Moreover, the equilibrium position obtained without imposition of symmetry (light blue diamond) roughly respects the symmetry of the system. 
This is due to the fact that the MSM configuration at each step of the relaxation is chosen randomly, acting on average according to the symmetry properties of the underlying crystal structure.

An important detail to take into account regards the size of the steps with which the atomic positions are updated during the relaxation, i.e. the magnitude of the parameter $\alpha$ in Eq. \ref{eq_xj+1}.
In the first stage of the process, the steps can be relatively large since the system needs to relax considerably; later, the size should be progressively reduced 
in order to sample adequately the region of interest, and avoid spurious local minima that are due to a particular MSM configuration where the atoms could get trapped. Nonetheless, we found that for Fe vacancy in PM bcc Fe, 
larger steps in the last part of the relaxations give just larger fluctuations around the same average value.
The parameter $\alpha$ depends on the software employed for the calculations: as an example, in VASP $\alpha$ is proportional to the POTIM-tag. 
In the case just shown, the value of POTIM started at 0.5 and then was decreased to 0.1. 

\begin{figure}
\begin{center}
\includegraphics[width=0.45\textwidth]{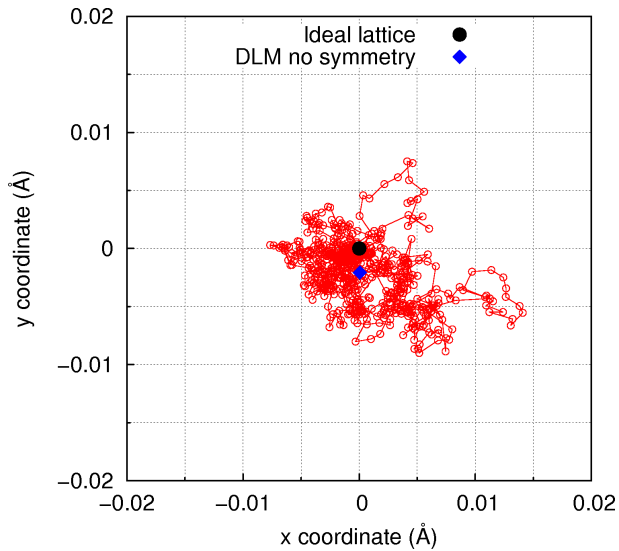}
\caption{x-y projection of the trajectory of one atom in the DLM-relaxation of defect free bcc Fe without imposition of symmetry. 
The red empty circles are the positions spanned by the atom during the relaxation; the black solid circle indicates the ideal lattice position, whereas the blue diamond is the resulting equilibrium DLM-relaxed position.
Note that the scale of the axes is smaller than in Fig.\ref{trajectory_1vbccFe.a}  \label{trajectory_dfbccFe}}
\end{center}
\end{figure}

In order to assess the accuracy of the present method without imposition of symmetry, we compare the position of an atom in defect free bcc Fe obtained with the DLM relaxation, to the ideal lattice site.
In Fig. \ref{trajectory_dfbccFe}, the trajectory of one atom during the relaxation is shown. 
In this case the FM-relaxed, DLM-relaxed with imposition of symmetry and the ideal lattice positions are coincident for symmetry reasons. 
As can be seen from Fig. \ref{trajectory_dfbccFe}, the trajectory during the relaxation evolves around the ideal lattice position on a scale much smaller than in the vacancy case. 
When calculating energies on these two geometries with the MSM method, the difference in energy is $<1$ meV/supercell, below the accuracy of DFT, probably due to the small difference between the positions and the flat energy landscape resulting from the DLM state. 
Regarding differences in DLM forces (related to corresponding atoms in the two geometries), the largest difference results to be $\sim 0.015$ eV/\angstrom, which is in the range of the statistical noise due to finite size of the sampling.

\section{Results \label{Results}}

\subsection{Fe vacancy in PM bcc Fe\label{1vbccFe}}

\begin{figure}%
\centering
\subfigure{\label{disp_1v_bccFe} \includegraphics[width=\columnwidth]{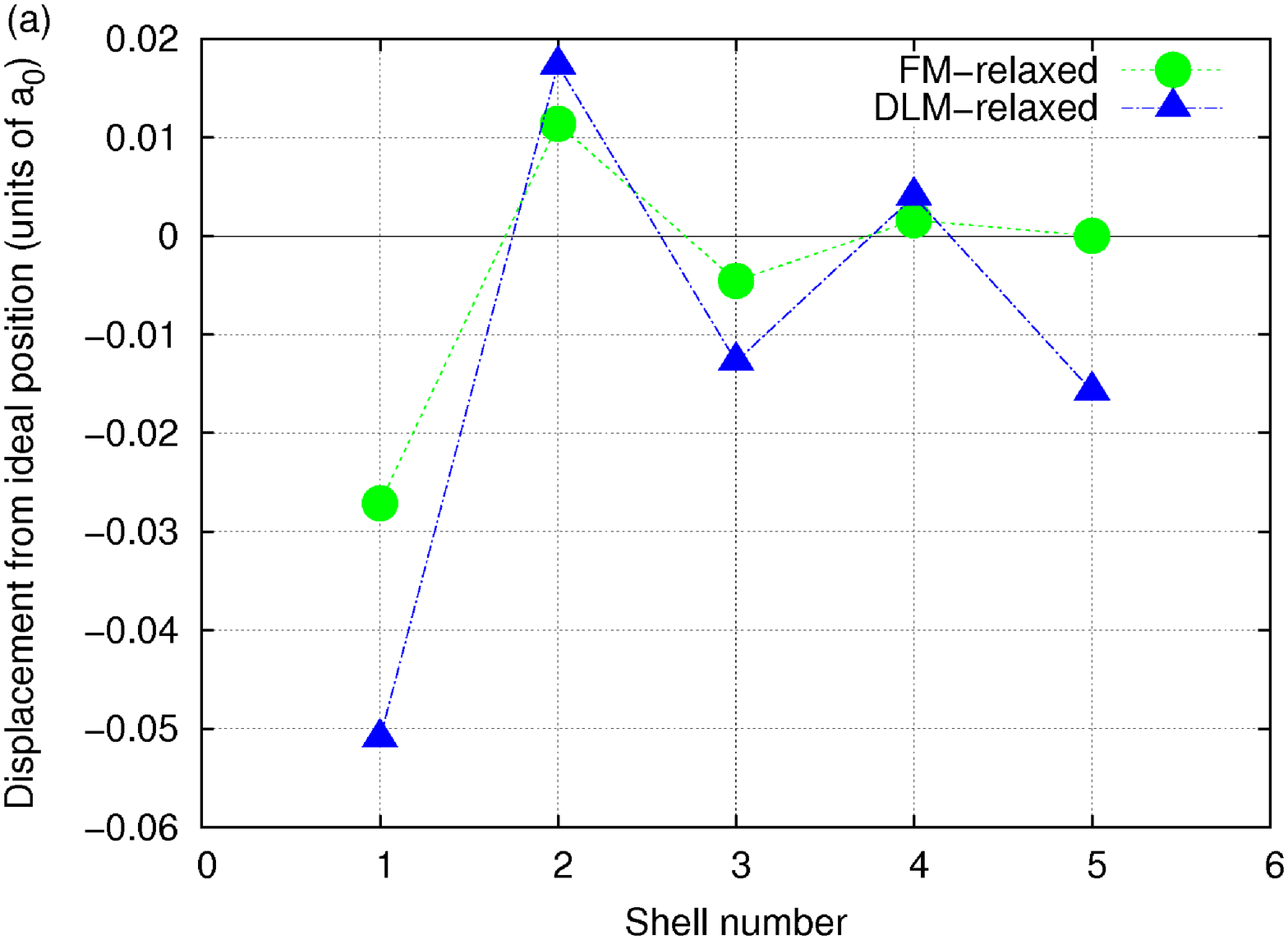}}
\subfigure{\label{avF_1v_bccFe} \includegraphics[width=\columnwidth]{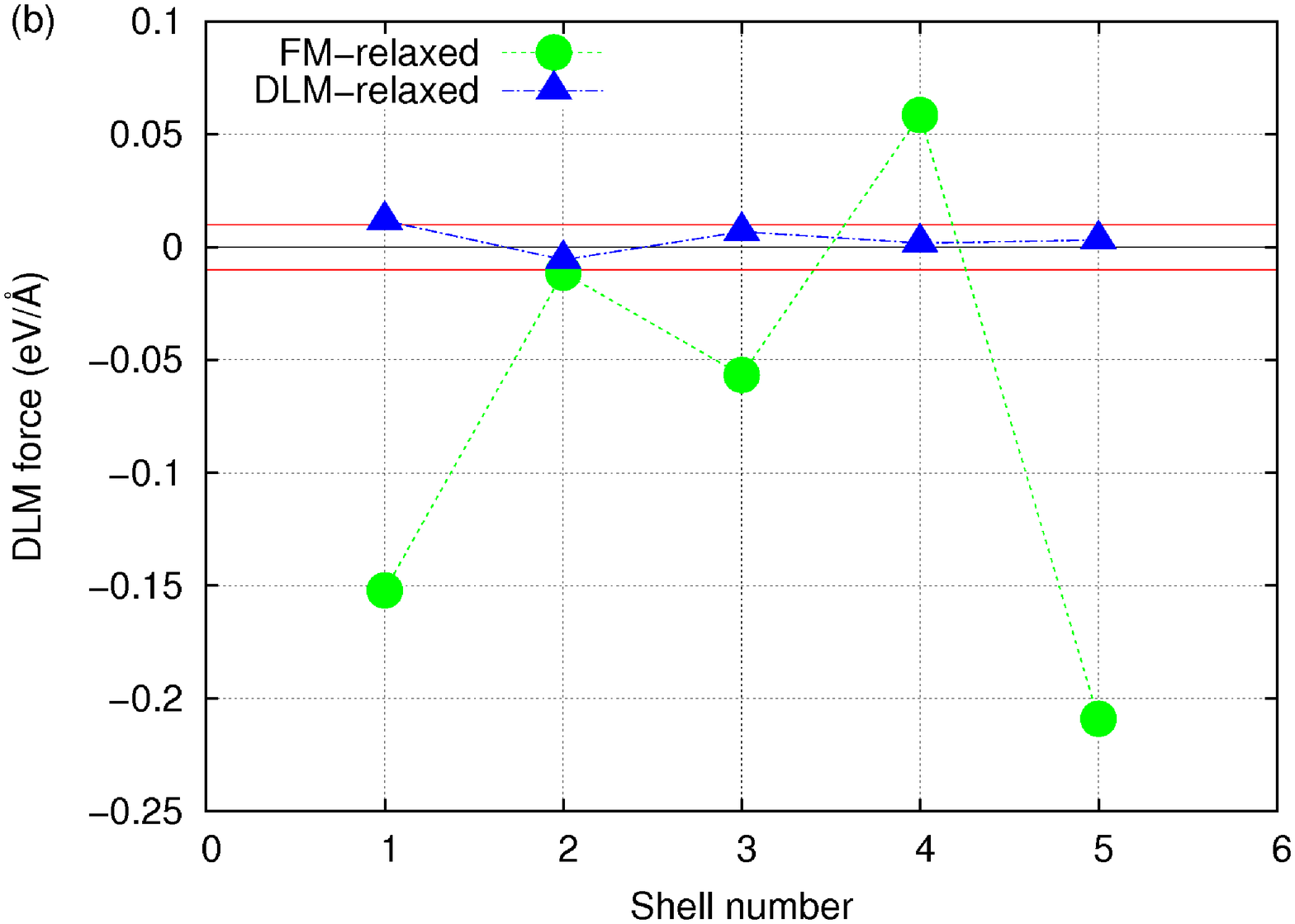}}
\caption{(a) Displacements from ideal lattice positions in units of lattice parameter $a_0$ and (b) DLM forces acting on atoms as a function of their distance to the vacancy for FM- and DLM-relaxed positions in bcc Fe with one vacancy. 
Negative values mean that the displacement (force) is directed towards the vacancy, positive is directed outwards. In (b), the red solid lines indicate a typical threshold for convergence of forces of 0.01 eV/\angstrom. 
\label{disp_avF_1v_bccFe}}
\end{figure}

The displacement of each independent atom from its ideal lattice position obtained from FM- and DLM-relaxation is shown in Fig. \ref{disp_1v_bccFe} as a function of the coordination shell.
The first nearest neighbors relax towards the vacancy of 0.14 \angstrom, which means that the distance from the vacancy reduces from 0.866 $a_0$ for the unrelaxed supercell to 0.815 $a_0$, where $a_0$ is the lattice parameter, 
which compares well with previous DMFT calculations \cite{DMFT_1vFe} (0.817 $a_0$).
It is obvious that for the nearest neighbors to the vacancy, the change of magnetic state affects importantly the atomic positions, since the interatomic forces become weaker consequently allowing larger relaxations; 
however, it is less obvious that for atoms in farther shells relaxations may be considerable as well. 
In this case, the atoms in the fifth shell are subject to a relaxation towards the vacancy of 0.017 $a_0$. 
This effect is a consequence of the relaxation of the nearest neighbors to the vacancy, since the atoms in the fifth shell are nearest neighbors to the atoms in the first shell along the line from the defect.

The  DLM forces on the atoms (Fig. \ref{avF_1v_bccFe}) in the DLM-relaxed positions are considerably smaller than in the FM-relaxed positions, below 0.01 eV/$\angstrom$  except for the nearest neighbors to the vacancy (0.014 eV/$\angstrom$). 
These small residual forces on the DLM-relaxed atoms have little effect on the energetics of the system, however they may lead to numerical problems in phonon calculations. 
The inability to reduce the forces arbitrarily close to zero is mainly due to statistical problems,  both during the relaxation and in the final calculations on the obtained relaxed positions.  
This is not a problem in the defect free case because the number of symmetry operations in the system are enough to ensure convergence (48 symmetry operations per atom in a 54-atoms supercell), 
whereas in the presence of a defect this number decreases consistently (48 symmetry operations for the atom with highest symmetry, 4-8 for the other atoms in a 53-atoms supercell) leading to lower statistical accuracy.

\begin{table}
\begin{center}
\caption{Vacancy formation energy for FM bcc Fe, and PM bcc Fe. PM bcc Fe is modelled in the present work with the DLM approach, and the formation energy is calculated on FM- and DLM-relaxed positions. 
In the DFT+DMFT investigation, the relaxed value was calculated minimizing the forces acting on the first two coordination shells. All values are in eV. \label{table_vac_form_energy}}
\begin{tabular}{@{\extracolsep{10pt}}cccc@{}}

\hline
\\[-1em]
& FM & \multicolumn{2}{c}{PM} \\
\cline{2-2} \cline{3-4}
\\[-1em]
& FM-relaxed & FM-relaxed & DLM-relaxed \\
\hline
This work & 2.20 & 1.70$\pm$ 0.06 & 1.61 $\pm$ 0.06 \\
DFT+DMFT \footnote{Delange et al. \onlinecite{DMFT_1vFe}} & 2.45 & 1.66 $\pm$ 0.15 & 1.56 $\pm$ 0.13 \\
DLM \footnote{Sandberg et al. \onlinecite{DLM_1vFe_Korzhavyi}} & 2.15 & 1.54 $\pm$ 0.16 & \\
SW \footnote{Ding et al. \onlinecite{SW_1vFe_Asta}} & 2.13 & 1.98 & \\
Experimental \footnote{Average of several experimental results (FM from Refs. \onlinecite{exp1vFe_I,exp1vFe_V,exp1vFe_IV}, PM from Refs. \onlinecite{exp1vFe_I,exp1vFe_II,exp1vFe_III,exp1vFe_IV,exp1vFe_V})} & 1.8 & \multicolumn{2}{c}{1.6} \\ 
\hline
\end{tabular}
\end{center}
\end{table}

The relaxations of the atomic positions in the DLM state have a clear effect on the vacancy formation energy. 
In Table \ref{table_vac_form_energy}, the present results are compared with previous theoretical calculations obtained with different methods and experimental measurements, for FM state and PM state with and without relaxation in the relevant magnetic state.
Here, the experimental value in the PM phase is the average of several experimental positron annihilation spectroscopy measurements\cite{exp1vFe_I,exp1vFe_II,exp1vFe_III,exp1vFe_IV,exp1vFe_V}. 
It should be noted that an exact agreement between theoretical formation energies and experimental measurements at elevated temperatures is not to be expected due to e.g. consequences of anharmonic vibrations \cite{PRX_Glensk}. 
We include them in the comparison to show the experimental trend resulting from change of magnetic state.

The DFT+DMFT \cite{DMFT_1vFe} and spin-wave (SW) \cite{SW_Ruban,SW_1vFe_Asta} results are calulated considering the experimental lattice parameter at the Curie temperature, 
whereas the present results and the DLM results in Ref. \onlinecite{DLM_1vFe_Korzhavyi} are calculated with the theoretical 0 K lattice parameter. 
In general, one can expect that the employment of the expanded lattice constant leads to a lower formation energy. 
The results from the SW method give considerably higher values of the formation energy compared to the other theoretical results, even though the expanded lattice parameter was employed. 
Although the SW method has a firm theoretical basis,  its overestimation could be due to the fact that only ordered configurations of the magnetic moments are employed there.
The DLM results from Sandberg et al. [\onlinecite{DLM_1vFe_Korzhavyi}] give a low vacancy formation energy, despite the fact that relaxation is done only with FM moments. 
This underestimation is probably due to the fact that in Ref. \onlinecite{DLM_1vFe_Korzhavyi} many MSM configurations had been neglected because of spin-flips during the electronic calculations, so that a statistically biased subset of configurations were taken into account.
In the present investigation, we have carried out calculations in a constrained framework also to avoid this technical problem, and we do obtain more consistent results.
Our result compares well with the DMFT result from Ref. \onlinecite{DMFT_1vFe}, even though the models to describe the PM phase are so different.
Part of the small difference between the formation energy calculated with the two different methods is probably due to the different lattice parameters employed. 
If we performed a calculation with the present method at the same lattice parameter as in [\onlinecite{DMFT_1vFe}], we would expect a lower formation energy due to two reason.
The first is that in Ref. \onlinecite{DMFT_1vFe}, only the first two coordination shells were allowed to relax, wheras we allow for full relaxation leading to larger relaxation energies and more consistent results.
The second is that our DLM magnetic state corresponds to a maximally disordered magnetic state, whereas the DMFT one correspond to a mean field approximation of finite temperature: for this reason, we can expect that our formation energy would be even smaller.
In order to obtain fully comparable results with experiments, the same workflow as described in [\onlinecite{PRL_Kormann_2014}] should be employed, i.e., 
 we should perform the DLM relaxation at the experimental lattice parameter, calculate on the resulting positions the vacancy formation energy both in the FM and the DLM state, 
and finally perform a weighted average of the two values according to the degree of short-range order at the given temperature, but this is beyond the scope of the present paper. 
Additionally, also the effect of lattice vibrations should be taken into account in order to be fully consistent with the experimental situtation.

\subsection{C interstitial in octahedral position in PM bcc Fe \label{octCbccFe}}

\begin{figure}%
\centering
\subfigure{\label{disp_octC_bccFe} \includegraphics[width=\columnwidth]{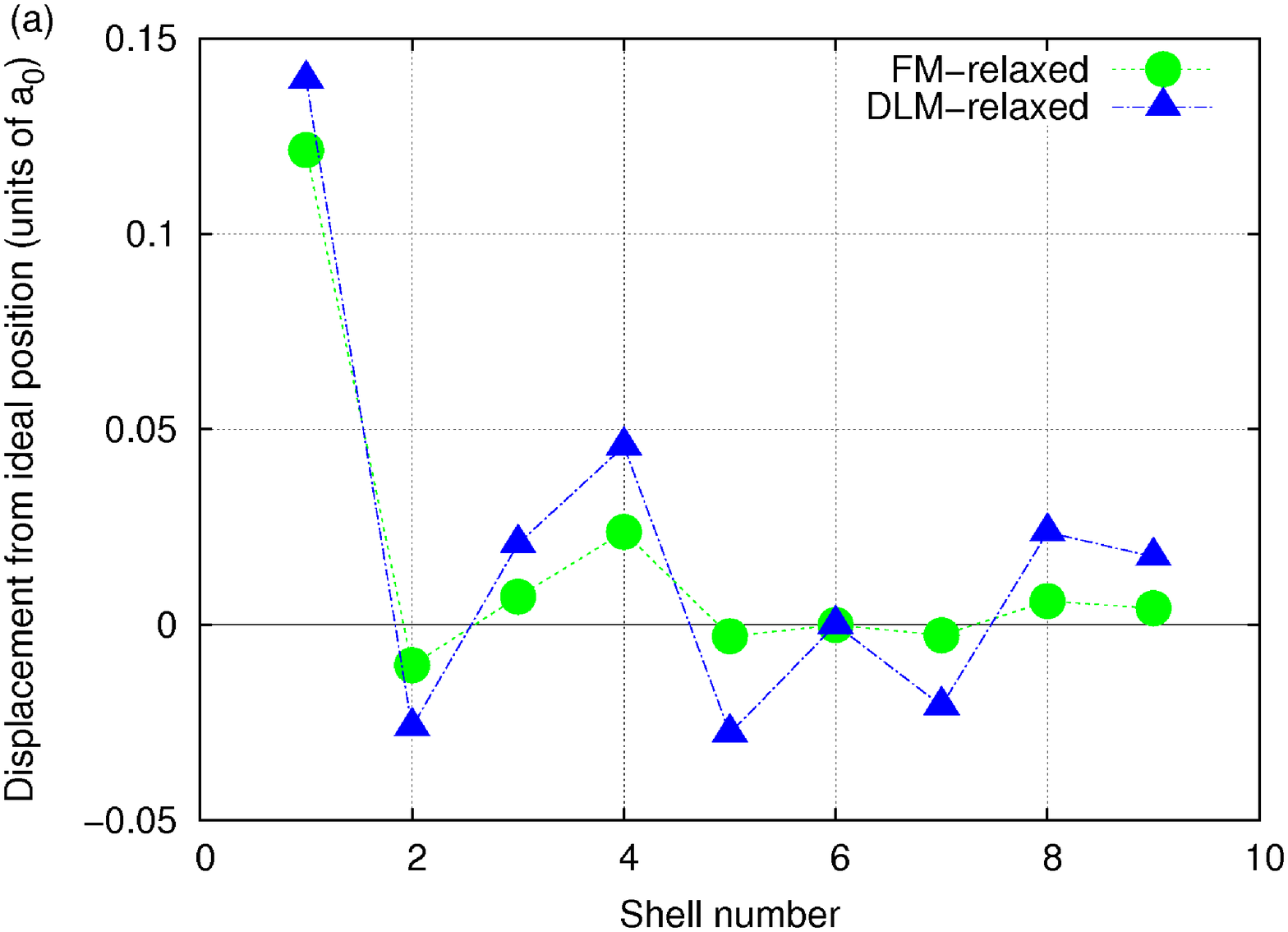}}
\subfigure{\label{avF_octC_bccFe} \includegraphics[width=\columnwidth]{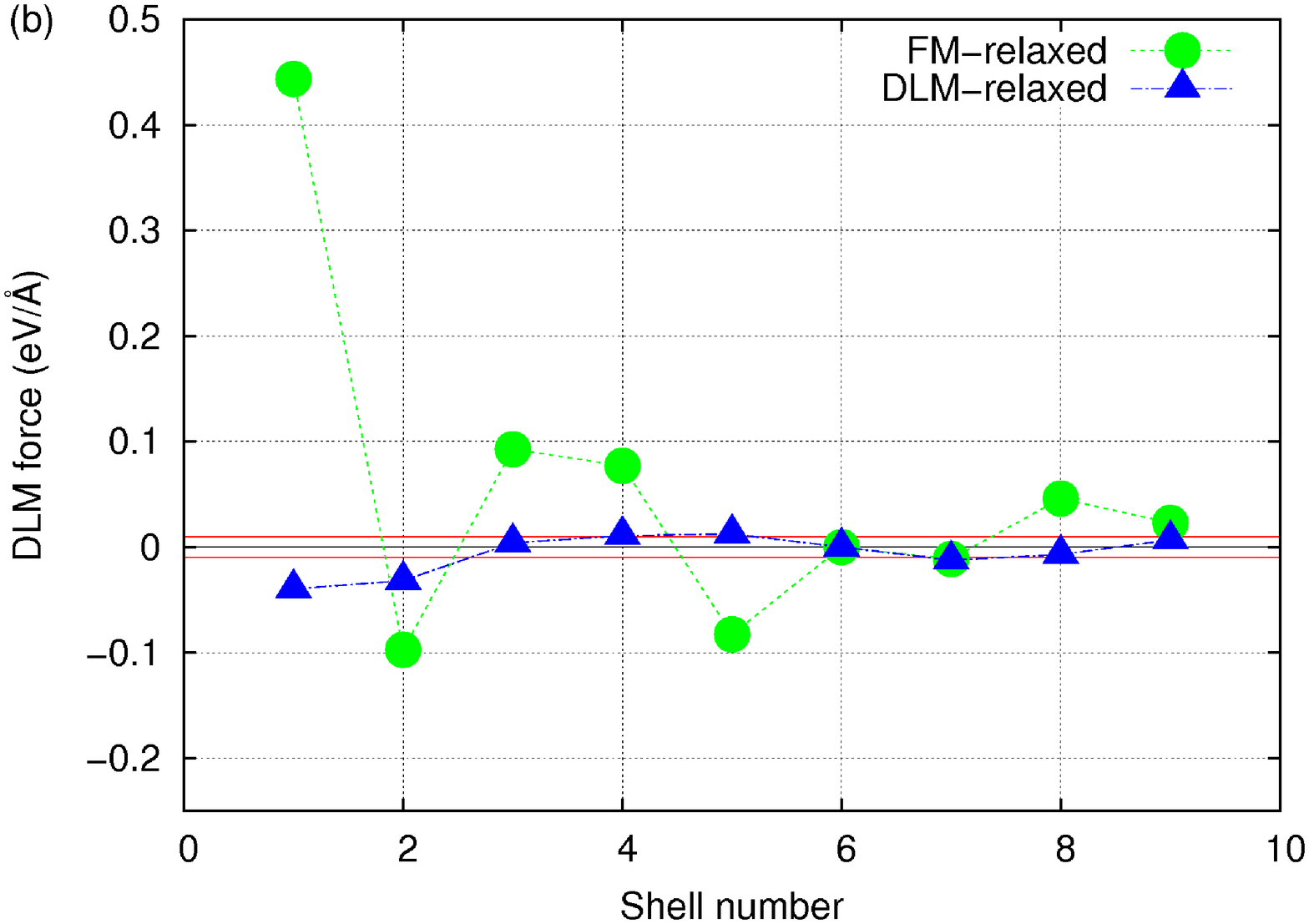}}
\caption{(a) Displacements from ideal lattice positions in units of lattice parameter $a_0$ and (b) DLM forces as a function of the coordination shell to the C interstitial for FM- and DLM-relaxed positions in bcc Fe with one C interstitial in octahedral position. 
Negative values mean that the displacement (force) is directed towards the interstitial, positive is directed outwards. In (b), the red solid lines indicate a typical threshold for convergence of forces of 0.01 eV/\angstrom. 
\label{disp_avF_octC_bccFe}}
\end{figure}

The carbon atom in interstitial position induces a change of symmetry from cubic to tetragonal, so that the sites independent from each other by symmetry are different from the ones in the vacancy case. 
In Fig. \ref{disp_octC_bccFe} and Fig. \ref{avF_octC_bccFe} the displacement from ideal lattice positions and the residual DLM forces of the FM- and DLM-relaxed geometries are respectively shown as a function of the coordination shell,
where positive (negative) displacement means that the atom goes away from (towards) the interstitial, and similarly for the DLM forces.
The displacements from ideal lattice positions are on average larger than in the case of the vacancy, for both FM- and DLM-relaxed geometries. 
Also here in the DLM-relaxation we find consistent differences with respect to the FM-relaxed positions, both for shells close to the defect and farther away. 
Non-intuitive considerable displacement for shells far from the interstitial atom (4th and 5th shells) due to the rearrangement of the atoms in shells closer to the C atom are observed also in this case.
The residual DLM forces calculated on DLM-relaxed positions are below 0.01 eV/$\angstrom$ for all the shells except for the first two. 
The first shell displays a residual force which goes against the displacement obtained from DLM-relaxation, whereas for the second shell the displacement seems to be underestimated. 

In Table \ref{table_C_interstitials_form_energies}, formation energy of the C interstitial in octahedral position is reported for FM and DLM states, and for this last state both results on FM- and DLM-relaxed positions are shown. 
The supercell size is important in these calculations, at least in the FM state, as it can be noted from the difference in C formation energy of 0.1 eV when calculated with a 3x3x3 or a 4x4x4 supercell.
This consistent difference is due to the fact that the larger supercell can accomodate the strain introduced by the interstitial atom on a larger region of the system, leading to a lower formation energy.
The two different supercell sizes are representative of different concentrations of carbon in iron (0.7 at\% and 2 at\% for the larger and smaller supercells, respectively). 
The solubility of C in the low-temperature bcc phase of Fe is at most $\sim$0.1 at\% \cite{FeC_phase_diagram}, but other phases in steels related to ferrite with larger concentrations of carbon, 
e.g. martensite, are of high technological relevance so that it is important to study also these high concentrations.
For this reason, we choose to investigate the smaller supercell in the DLM state.
In addition, the qualitative behavior is not expected to change for the two different supercell sizes, but only the quantitative results could be affected.
The DLM-relaxation of the octahedral C in the 3x3x3 supercell shows a decrease in formation energy of 0.18 eV compared to the FM-relaxed positions, which is larger than the decrease due to the change of magnetic state from FM to DLM (0.09 eV).

\begin{table}
\begin{center}
\caption{Interstitial formation energy for C in the octahedral position in FM and PM bcc Fe. PM bcc Fe is modelled in the present work with the DLM approach. As a comparison, the formation energy calculated with a 4x4x4 supercell in the FM state is also shown. All values are in eV. \label{table_C_interstitials_form_energies}}
\begin{tabular}{@{\extracolsep{10pt}}cccc@{}}
\hline
\\[-1em]
& FM & \multicolumn{2}{c}{PM} \\
\cline{2-2} \cline{3-4}
\\[-1em]
& FM-relaxed & FM-relaxed & DLM-relaxed \\
\hline
3x3x3 supercell & 0.68 & 0.59 $\pm$ 0.07 & 0.41 $\pm$ 0.06 \\
4x4x4 supercell & 0.58 & - & -  \\ 
\hline
\end{tabular}
\end{center}
\end{table}

Up to our knowledge, no theoretical investigation from first principles of C interstitials in bcc Fe in the PM phase has ever been performed with an accurate description of the magnetic state, so that our results cannot be compared with other theoretical calculations. 
Moreover, the present DLM description of the PM phase can be considered as a good description of the magnetism of Fe in the $\delta$-phase, so that the present results should be compared with experimental measurements in this high temperature phase in order to be consistent; 
unfortunately, the $\delta$-phase has not been studied as much as the lower temperature phases of the Fe-C phase diagram, so experimental estimations of the formation energies are not available to assess our results. 
Nonetheless, a decrease in formation energy is reasonable going from the FM to the PM state.

As a final remark, we report that a relaxation of bcc Fe in the DLM state with a C interstitial in octahedral position  without imposition of symmetry leads to a formation energy 30 meV lower than in the symmetrized case. 
An accurate calculation of the formation energy should investigate more in detail this effect, going also towards the dilute limit.

\subsection{PM bcc Fe$_{1-x}$Cr$_{x}$ alloy \label{bccFeCr}}

\begin{figure}
\begin{center}
\includegraphics[width=\columnwidth]{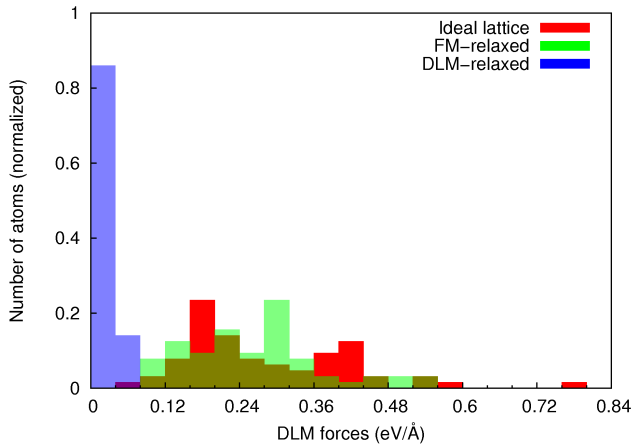}
\caption{Average forces in DLM bcc Fe$_{0.5}$Cr$_{0.5}$ alloy calculated on ideal lattice (red), FM-relaxed (green), and DLM-relaxed (blue) positions. The area of each set is normalized to one. \label{avforce_bccFeCr}}
\end{center}
\end{figure}

In random alloys, lattice symmetry is on average retained, but on the level of individual atoms, the symmetry is completely broken: thus, in the calculations of PM bcc Fe$_{1-x}$Cr$_{x}$ alloys, one cannot rely on the increased statistics obtained from symmetrization. 
Also in this case the DLM forces on the atoms are substantially reduced when the DLM-relaxed positions are employed, compared to the FM-relaxed and ideal lattice structures, as shown in the case of bcc Fe$_{0.5}$Cr$_{0.5}$ alloy in Fig. \ref{avforce_bccFeCr}.
It is interesting to notice here that the FM-relaxed positions do not give clearly better DLM forces compared to the ideal lattice ones, stressing the importance of the present method for DLM-relaxations in this alloy.

In these DLM calculations, the magnetic moments on the Cr atoms are fairly small ($\sim 0.1$ $\mu_B$), whereas the Fe atoms show larger moments in the range 1.5-2.2 $\mu_B$.

\begin{figure}%
\centering
\subfigure{\label{rdf_Cr-Cr_bccCrFe} \includegraphics[width=\columnwidth]{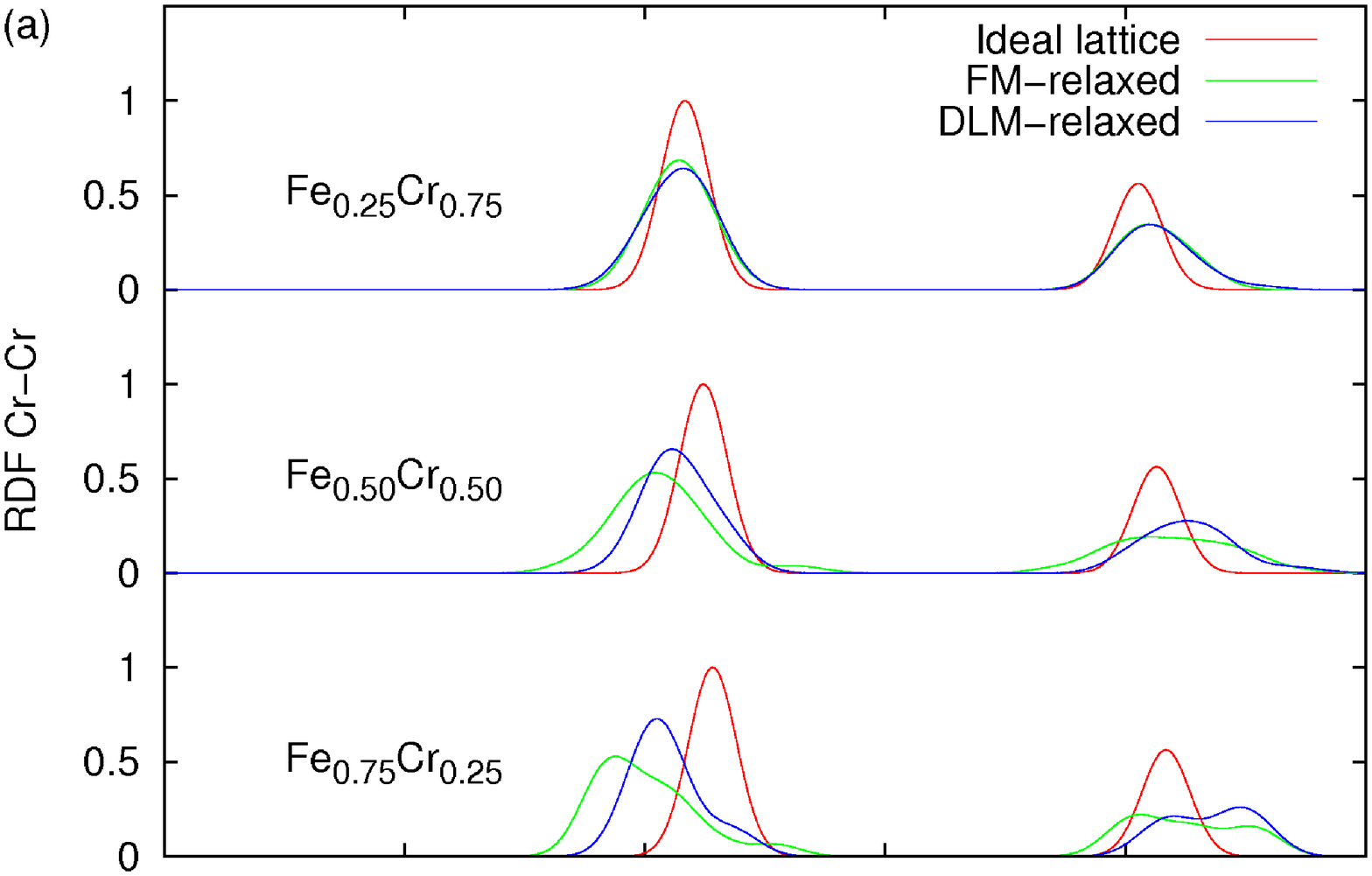}}
\subfigure{\label{rdf_Cr-Fe_bccCrFe} \includegraphics[width=\columnwidth]{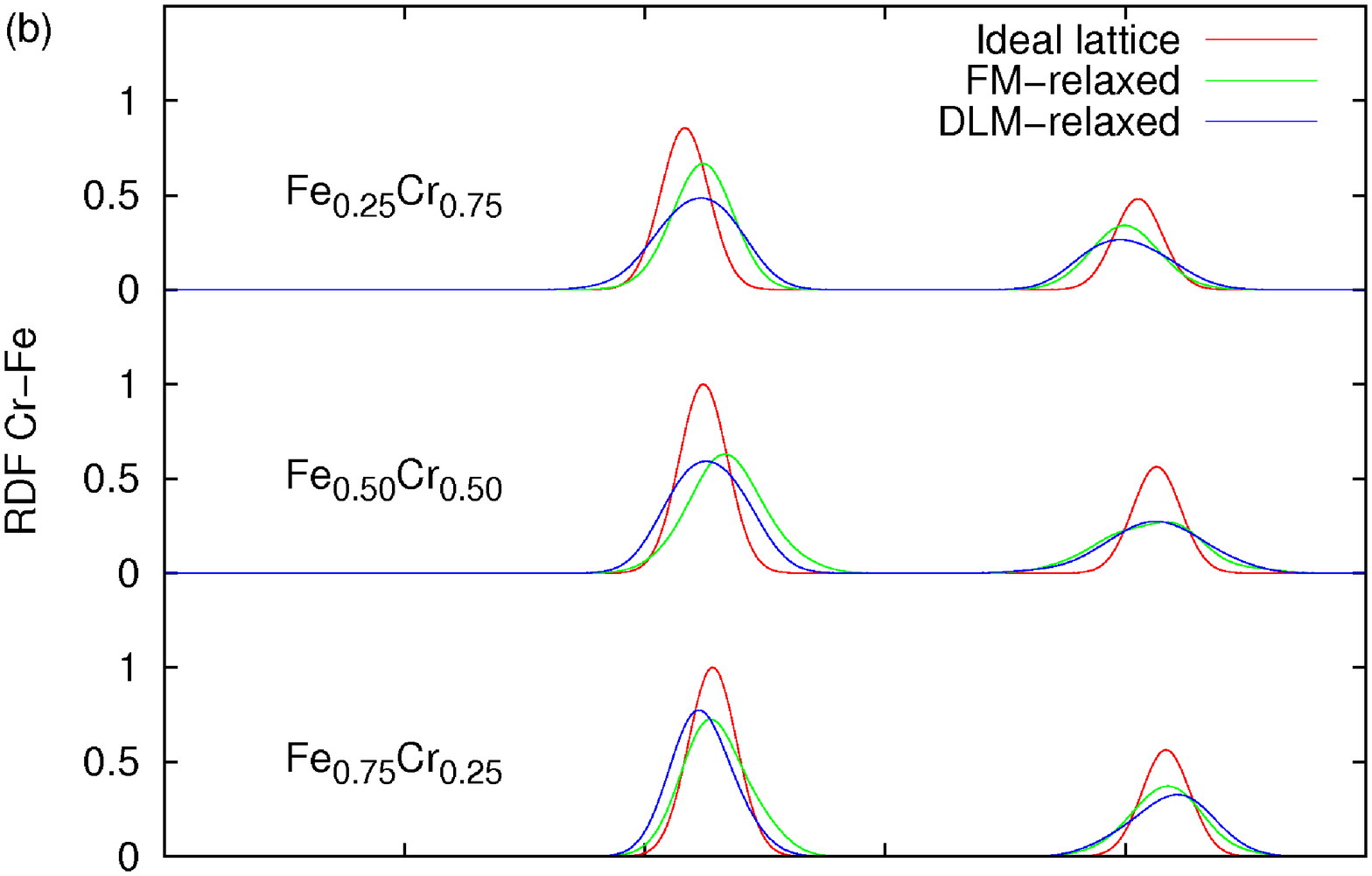}}
\subfigure{\label{rdf_Fe-Fe_bccCrFe} \includegraphics[width=\columnwidth]{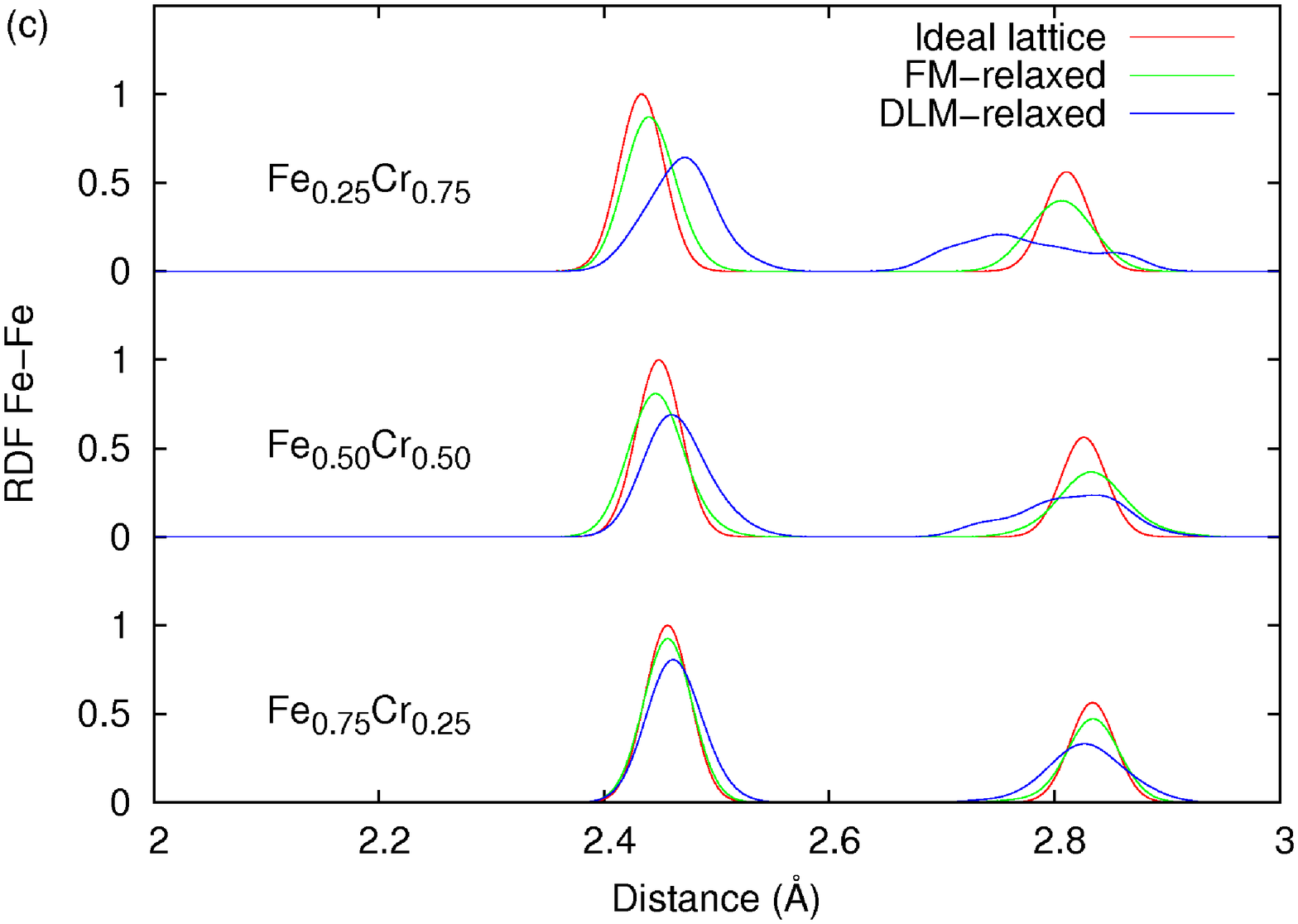}}
\caption{Radial distribution functions for bcc Fe$_{1-x}$Cr$_{x}$ alloys per bond type. (a) Cr-Cr bond; (b) Fe-Cr bond; (c) Fe-Fe bond. All the curves are rescaled with the height of the highest peak equal to 1. 
\label{rdf_bccCrFe}}
\end{figure}

In Fig. \ref{rdf_bccCrFe}, the radial distribution function (RDF) for every type of bond (\ref{rdf_Cr-Cr_bccCrFe} Cr-Cr, \ref{rdf_Cr-Fe_bccCrFe} Fe-Cr, and \ref{rdf_Fe-Fe_bccCrFe} Fe-Fe) is shown for the first two coordination shells at every composition. 
In general, the peaks become broader than in the ideal lattice structure. 
If one focuses on the Cr-Cr first coordination shell, it can be noticed that going from low to high Cr content, the peak becomes more similar to the ideal one, as it could be expected since for pure bcc Cr the ideal lattice positions are also the equilibrium ones.
The Fe-Fe distance, Fig. \ref{rdf_Fe-Fe_bccCrFe}, in first coordination shell follows a similar trend, i.e. for increasing Fe content the nearest neighbors distance becomes more similar to ideal lattice. 
The difference between them is that for lower Cr content, the Cr-Cr distance becomes smaller than in the ideal lattice structure, whereas the Fe-Fe distance increases for decreasing Fe content.

It is interesting to see also the different behavior of the FM- and DLM-relaxed nearest neighbors peaks. The RDFs of the Fe-Fe bond FM-relaxed is closer to the ideal lattice peak than the DLM-relaxed for every composition;
the Cr-Cr bond, in contrast, is farther from ideal position for FM-relaxed than DLM-relaxed, especially for low Cr content.

Regarding the second nearest neighbors, the peaks become even broader, up to the point that for Fe$_{0.75}$Cr$_{0.25}$ there seems to be two peaks. 
This could be an effect of the local environment, and the appearence of a second peak could be due only to the small size of the supercell; however, a more detailed investigation with larger SQS cells would be needed to clarify if there is a simple peak broadening or indeed a developed two-peak feature.

\begin{figure}
\includegraphics[width=\columnwidth]{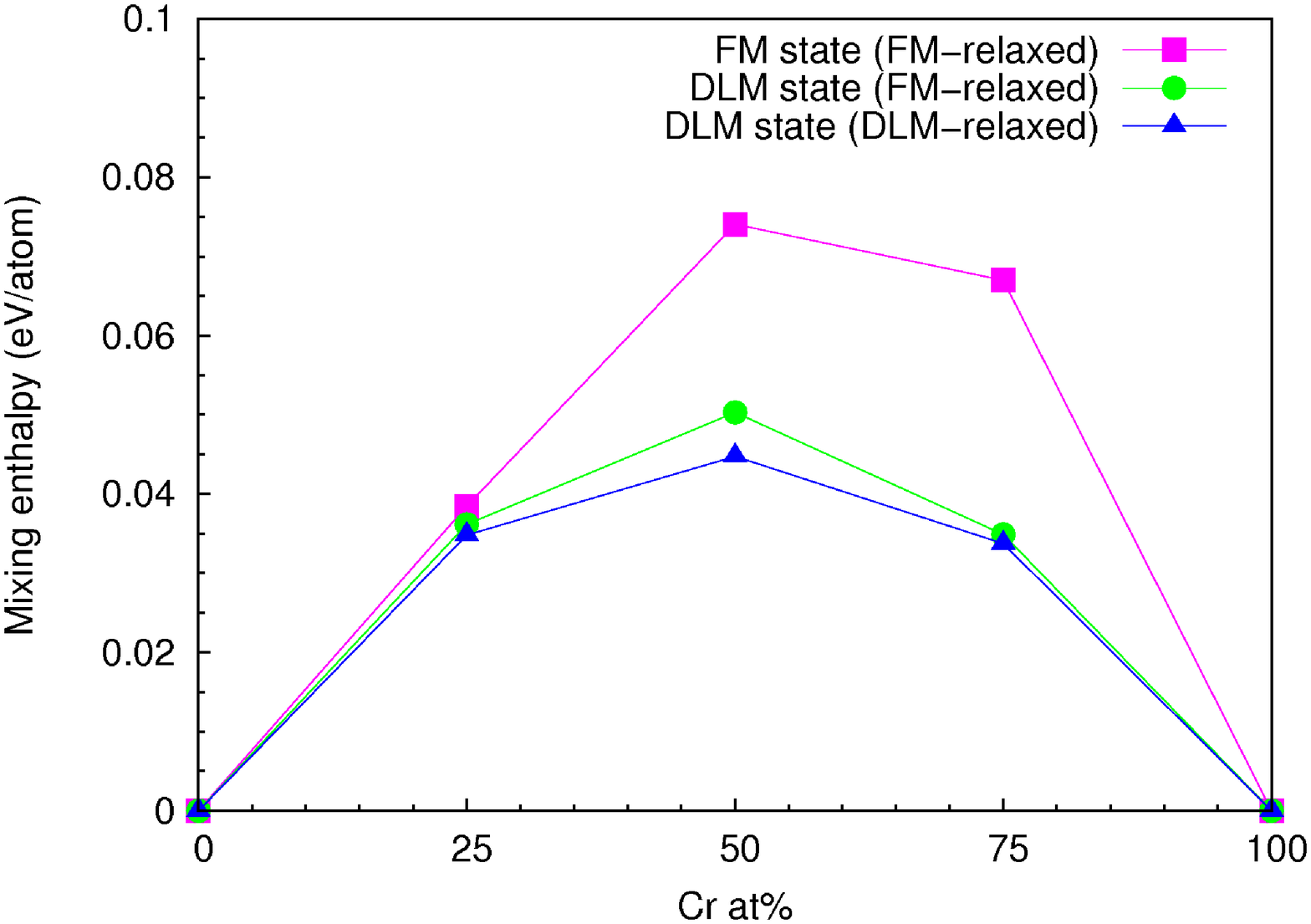}
\caption{Mixing enthalpy of bcc Fe$_{1-x}$Cr$_{x}$ alloys for x=0.25, 0.50, and 0.75 in FM state on FM-relaxed positions (purple squares) and in DLM state on FM-relaxed positions (green circles) and DLM-relaxed positions (blue triangles). 
The lines are just guidelines for the eyes. \label{Mixing_enthalpy_bccFeCr}}
\end{figure}

The mixing enthalpy of bcc Fe$_{1-x}$Cr$_{x}$ alloys is known \cite{bccCrFe_Igor1,bccCrFe_Igor2} to show a qualitative difference between FM and DLM state. 
FM alloys show, indeed, a region for small Cr concentrations where the mixing of Fe and Cr is favorable, and for larger concentrations the usual solubility gap is present. The DLM state removes this small region of solubility and in general decreases the magnitude of the positive mixing enthalpy.
In Fig. \ref{Mixing_enthalpy_bccFeCr} the mixing enthalpy of bcc Fe$_{1-x}$Cr$_x$ is shown for the composition considered in the present work, where the lines are just a guide for the eyes. 
The reference states are here DLM bcc Fe and nonmagnetic bcc Cr for the alloys in the DLM state, the latter being the magnetic result when we start with a DLM bcc Cr. Analogously, for the alloys in the FM state, FM bcc Fe and nonmagnetic bcc Cr are the reference states.
The small region of solubility for the FM alloy is not visible here because it occurs at concentrations lower than x=0.10\cite{bccCrFe_Igor2}; however, its presence can be guessed from the strongly asymmetric behavior of the FM curve.
In the DLM magnetic state, the employment of the DLM-relaxed positions does not change qualitatively the curve compared to the FM-relaxed positions, rather it introduces an energy correction to the mixing enthalpy. 
The correction is largest at x=0.50 ($\sim 5$ meV/atom or about 10\%), whereas for the other two compositions the effect is smaller ($\sim 1$ meV/atom). This difference is due to the fact that the closer the alloy is to the pure elements, the smaller are the relaxation energies per atom.

\section{Conclusions \label{Conclusions}}

In this work we have developed a method that allows to perform local lattice relaxation of magnetic materials in the PM phase from first principles within the disordered local moments model in an adiabatic fast-magnetism framework. 
The method is easy to implement with any $ab$ $initio$ code that allows to calculate reliably interatomic forces. 
We employ constrained noncollinear magnetic moments in this work because they give a more correct description of the real paramagnetic state; 
moreover, we observe a difference of the DLM energy for bcc Fe compared to collinear arrangement of the atomic magnetic moments (ca. 3 meV/atom of difference).
Noncollinear moments also lead to smaller standard deviations in properties such as atomic forces, enabling a more accurate relaxation of the systems.
We first test the present method on the case of a Fe vacancy in PM bcc Fe, a well studied system, and we obtain a vacancy formation energy of 1.60 eV, in good agreement with recent DFT+DMFT results [\onlinecite{DMFT_1vFe}] and experimental measurements.
The C interstitial in octahedral position is then investigated, and the formation energy of this defect in the DLM state is 0.41 eV, 
where the difference from the value in the FM state is given in large part by the DLM-relaxed positions rather than the change of magnetic state itself.
Finally, we calculate the mixing enthalpy for bcc Fe$_{1-x}$Cr$_{x}$ alloys for x= 0.25, 0.50, and 0.75; here, the DLM-relaxed positions lead to a reduction of the mixing enthalpy of 5 meV/atom for x=0.50 ($\sim 10 \%$).
All systems relaxed with the present method show lower DLM forces than the ferromagnetically relaxed counterparts.

\section*{Acknowledgments}
This research was carried out using computational resources provided by the National Supercomputer Centre (NSC) in Link\"oping (Gamma supercomputer) and the Swedish National Infrastructure for Computing (SNIC): Triolith Cluster located at NSC.
B.A. acknowledges financial support by the Swedish Research Council (VR) through the International Career Grant No. 2014-6336 and by Marie Sklodowska Curie Actions, Cofund, Project INCA 600398, the Swedish Government Strategic Research Area in Materials Science on Functional Materials at Link\"oping University (Faculty Grant SFOMatLiU No 2009 00971), as well as support from the Swedish Foundation for Strategic Research through the Future Research Leaders 6 program. Tilmann Hickel is acknowledged for useful discussions.

\end{document}